\documentstyle[twocolumn,epsbox,subfigure]{IEEEtran}
\def\BibTeX{{\rm B\kern-.05em{\sc i\kern-.025em b}\kern-.08em
    T\kern-.1667em\lower.7ex\hbox{E}\kern-.125emX}}

\begin{document}


\pagestyle{empty}

\newcommand{\boldvec}[1]{\mbox{\boldmath $#1$}}
\newcommand{\alt}{\displaystyle{\mathop{<}_{\sim}}}

\title{\LARGE \bf 
Influence of Small Impurities on Low-Energy Electron Dynamics 
in Two-Dimensional Microscopic Bounded Region
} 

\author{
{\bf T. Shigehara}$^{1}$, 
{\bf M. Yokoyama}$^{1}$, 
{\bf H. Mizoguchi}$^{1}$, 
{\bf T. Mishima}$^{1}$ 
{\bf and Taksu Cheon}$^{2}$
\normalsize
$^{1}$Department of Information and Computer Sciences, Saitama University \\
Shimo-Okubo 255, Urawa, Saitama 338-8570, JAPAN \\
Phone: +81-48-858-9035,
Fax: +81-48-858-3716,
E-MAIL : sigehara@ics.saitama-u.ac.jp \\[0.5ex]
$^{2}$Laboratory of Physics, Kochi University of Technology \\
Tosa Yamada, Kochi 782-8502, JAPAN \\
Phone: +81-8875-7-2302,
Fax: +81-8875-7-2320,
E-MAIL: cheon@mech.kochi-tech.ac.jp
}
\date{}


\maketitle

\begin{abstract}
In order to give some insight into a role of small impurities 
on the electron motion in microscopic devices, 
we examine from a general viewpoint, the effect of small obstacles 
on a particle motion at low energy inside microscopic bounded regions. 
It will be shown that the obstacles disturb the electron motion 
only if they are weakly attractive.  
\end{abstract}

\setcounter{page}{1}
\setcounter{figure}{0}
\setcounter{table}{0}

\section{Introduction} 

The recent progress of microscopic technology 
makes it possible to construct extremely pure structures 
which are expected to be a main component of 
promising quantum devices such as a single-electron memory.  
However, real systems are not free from small impurities which 
might possibly affect the electron motion inside. 
It is an urgent problem to reveal the condition under which 
the wave function of the electron is substantially distorted 
by the small impurities. 
This is exactly our subject in this paper. 

We here restrict ourselves to the case of spatial dimension two. 
We begin in Sect.2 with a simple but exactly solvable 
model where a pointlike particle with mass $M$ freely moves in a 
two-dimensional bounded region which contains a pointlike scatterer 
inside. 
It is a well-known fact that the Dirac's delta potential does not work 
in quantum mechanics in spatial dimension two. 
Based on the self-adjoint extension of a symmetric operator in 
functional analysis, however, we derive a suitable transition 
matrix for the system with a pointlike interaction. 
By examining the general feature of the eigenvalue equation 
for the system, we deduce the general condition 
under which the eigenfunctions are substantially 
affected by the pointlike scatterer. 
In Sect.3, we consider a quantum-mechanical one-body problem with 
the potential which has constant strength $U_1$ in 
a small but finite region of size $\Omega$. 
The potential is expected to behave as pointlike at low energy 
where the electron wavelength is much larger than 
the range of the potential.  
Thus we can apply the findings for pointlike obstacles to such cases. 
Keeping the area $\Omega$ small but finite-size, 
we show that the electron motion at the energy $\omega$ is 
substantially distorted by the potential under the condition 
\begin{eqnarray}
\label{eq:1}
\left|\frac{1}{U_1M\Omega}-\frac{1}{2\pi}\ln(\omega M\Omega)\right|
\alt  \frac{\pi}{4}.   
\end{eqnarray}
This indicates that the small impurities 
influence the low-energy electron wave function 
only if they are weakly attractive.   
The validity of our conjecture is confirmed by numerical experiments. 
The current work is summarized in Sect.4. 

\section{The Case of Point Impurity}

We first consider a quantum point particle of mass $M$ 
moving freely in a two-dimensional bounded region $S$. 
Let us denote the area of $S$ by the same symbol. 
We impose the Dirichlet boundary condition so that wave functions 
vanish on the boundary of $S$. 
The eigenvalues and the corresponding normalized eigenfunctions 
are denoted by $E_n$ and $\varphi_n(\boldvec{x})$ respectively; 
\begin{eqnarray}
\label{eq:2}
H_0 \varphi _n(\boldvec{x}) \equiv -{\nabla^2 \over {2M}} 
\varphi _n(\boldvec{x})= E _n \varphi _n(\boldvec{x}). 
\end{eqnarray}
The Green's function of the kinetic operator $H_0$ is written as 
\begin{eqnarray}
\label{eq:3}
G^{(0)}(\boldvec{x},\boldvec{y};\omega ) = 
\sum\limits_{n=1}^\infty 
{{{\varphi _n(\boldvec{x})\varphi _n(\boldvec{y})} 
\over {\omega -E_n}}}, 
\end{eqnarray}
where $\omega$ is the energy variable. 
The average level density of the system is given by 
$\displaystyle \rho_{av} = MS/2\pi$, 
which is energy-independent. 
We now place a single point impurity at 
$\boldvec{x}_1$ in the region $S$. 
The most naive manner for this purpose is to 
define the impurity by using the Dirac's $\delta$ function of strength $v_1$;  
\begin{eqnarray}
\label{eq:4}
H=H_0+ v_1 \delta(\boldvec{x}-\boldvec{x}_1).
\end{eqnarray}
However, the Hamiltonian $H$ is not mathematically sound. 
This can be seen from the eigenvalue equation of $H$, 
which is reduced to 
\begin{equation}
\label{eq:5}
\sum_{n=1}^{\infty} 
\frac{\varphi_{n}(\boldvec{x}_1)^2}{\omega-E_n} = v_1^{-1}.
\end{equation}
Since the average level density $\rho_{av}$ is constant 
in spatial dimension two, 
the infinite series does not converge. 

One of the general schemes to handle the divergence  
is based on the self-adjoint extension theory of 
functional analysis \cite{AG88}--\cite{RS75}. 
We first restrict the domain of $H_0$, say $D(H_0)$, to the functions 
which vanish at the location of the point impurity; 
$\displaystyle H_{\boldvec{x}_1} =  -\nabla^2 /2M$,  
$\displaystyle 
D(H_{\boldvec{x}_1}) =  
\{ \psi \in D(H_0) \vert 
\psi (\boldvec{x}_1)=0  \}$.  
By using integration by parts, 
it is easy to prove that 
the operator $H_{\boldvec{x}_1}$ is symmetric (Hermitian). 
But it is not self-adjoint. 
Indeed, the eigenvalue equation 
$H_{\boldvec{x}_1}^* \psi_{\omega}=\omega\psi_{\omega}$  
for the adjoint of $H_{\boldvec{x}_1}$  
has a solution for $Im \ \omega \neq 0$ \cite{ZO80}; 
$ \displaystyle 
\psi_{\omega}(\boldvec{x}) =  
G^{(0)}(\boldvec{x},\boldvec{x}_1;\omega )$.  
Since the deficiency indices of 
$H_{\boldvec{x}_1}$ are $(1,1)$, 
$H_{\boldvec{x}_1}$ has one-parameter family of 
self-adjoint extensions $H_{\theta_1}$ ($0 \le \theta_1 < 2\pi$) 
which is regarded as the proper Hamiltonian for the system with 
a point impurity at $\boldvec{x}_1$. 
Following Zorbas \cite{ZO80}, 
we can write down 
the Green's function for $H_{\theta_1}$ as 
\begin{eqnarray}
\label{eq:6}
\lefteqn{
G_{\theta_1}(\boldvec{x},\boldvec{y};\omega) = 
G^{(0)}(\boldvec{x},\boldvec{y};\omega) 
} \nonumber \\
&& + G^{(0)}(\boldvec{x},\boldvec{x}_{1};\omega)
T_{\theta_1}(\omega)
G^{(0)}(\boldvec{x}_1,\boldvec{y};\omega),  
\end{eqnarray}
where the transition matrix $T_{\theta_1}$ is calculated by 
\begin{eqnarray}
\label{eq:7}
\hspace*{-5ex}
T_{\theta_1}(\omega) =\frac{1-e^{i\theta_1}}
{(\omega-i\Lambda) c_{i\Lambda}(\omega)-
e^{i\theta_1}(\omega+i\Lambda) c_{-i\Lambda}(\omega)}, 
\end{eqnarray}
with 
\begin{eqnarray}
\label{eq:8}
\hspace*{-5ex}
c_{\pm i\Lambda}(\omega)=
\int_{S}
G^{(0)}(\boldvec{x},\boldvec{x}_1;\omega)
G^{(0)}(\boldvec{x},\boldvec{x}_1;\pm i\Lambda)d\boldvec{x}. 
\end{eqnarray}
Here $\Lambda >0$ is an arbitrary scale mass. 
The eigenvalues of $H_{\theta_1}$ are determined by 
$T_{\theta_1}(\omega)^{-1}=0$, which is reduced to 
\vspace*{-3ex}
\begin{eqnarray}
\label{eq:9}
G(\omega)=\bar{v}_{1}^{-1}, 
\end{eqnarray}
%
where
\begin{eqnarray}
\label{eq:10}
G(\omega)=\sum_{n=1}^{\infty}
\varphi_{n}(\boldvec{x}_{1})^{2}
(\frac{1}{\omega-E_{n}}+\frac{E_{n}} {E_{n}^2+\Lambda^{2}}), 
\end{eqnarray}
\vspace*{-3ex}
\begin{eqnarray}
\label{eq:11}
\bar{v}_{1}^{-1}=\Lambda \cot \frac{\theta_1}{2}
\sum_{n=1}^{\infty}
\frac{\varphi_{n}(\boldvec{x}_1)^{2}}{E_{n}^2+\Lambda^{2}}. 
\end{eqnarray}
The constant $\bar{v}_{1}$ can be formally considered as the strength 
of the point impurity, 
the value of which ranges over the whole real number   
as $0\leq\theta_1 <2\pi$.  
On any interval $(E_m,E_{m+1})$, 
the function $G$ is monotonically decreasing, ranging 
over the whole real number. 
This means that the eigenvalue equation (\ref{eq:9}) has 
a single solution $\omega_m$ on each interval for any $\bar{v}_{1}$. 
The eigenfunction of $H_{\theta_1}$ corresponding to 
an eigenvalue $\omega_m$ is given by 
\begin{eqnarray}
\label{eq:12}
\hspace*{-3ex}
\psi_m (\boldvec{x}) \propto G^{(0)}(\boldvec{x},\boldvec{x}_1;\omega_m) 
= \sum_{n=1}^{\infty}
\frac{\varphi_n (\boldvec{x}_1)}{\omega_m - E_n} \varphi_n (\boldvec{x}). 
\end{eqnarray}

Based on the formulation described above, 
we deduce the condition for the appearance of the effect of 
point impurities on the particle motion. 
The first notice is that the average value of   
$\varphi_n (\boldvec{x}_1)^2$ among many $n$ 
is constant; 
$\left\langle \varphi_n (\boldvec{x}_1)^2 \right\rangle_n \simeq 1/S$. 
We thus recognize from (\ref{eq:12}) that 
if $\omega_m \simeq E_m$ (resp. $E_{m+1}$) for some $m$, 
then $\psi_m \simeq \varphi_m$ (resp. $\psi_m \simeq \varphi_{m+1}$).   
This implies that a point impurity distorts the wave function 
if the eigenvalue $\omega_m$ is located around the midpoint of 
the interval $(E_m,E_{m+1})$. 
For such $\omega_m$, 
the value of $G(\omega_m)$ can be estimated by using the principal integral,  
since the contributions on the summation of $G$ from 
the terms with $n \simeq m$ cancel each other.    
We thus realize that the point impurity of formal strength $\bar{v}_1$ 
causes the wave function mixing mainly 
in the eigenstate with an eigenvalue $\omega$ which satisfies 
\begin{eqnarray}
\label{eq:13}
\left| \bar{v}_1^{-1} -  \alpha \cdot
P  \int_{0}^{\infty} \left( 
\frac{1}{\omega-E}+\frac{E}{E^2+\Lambda^2} 
\right) dE \right| \alt \frac{\Delta}{2}  
\end{eqnarray}
with  
$\displaystyle 
\alpha = \left\langle \varphi_{n}(\boldvec{x}_{1})^{2} \right\rangle_n 
\rho_{av} = M/2\pi$,  
leading to 
\begin{eqnarray}
\label{eq:14}
\left| \bar{v}_1^{-1}- \frac{M}{2\pi}\ln\frac{\omega}{\Lambda} \right| 
\alt \frac{\Delta}{2}.   
\end{eqnarray}
The ``width'' $\Delta$ of the strong coupling region 
is estimated by considering the variance of $G$ linearized at 
$\omega = (E_m + E_{m+1})/2$ on the interval $(E_m, E_{m+1})$; 
\begin{eqnarray}
\label{eq:15} 
\Delta & \simeq & 
\left| G'(\omega) \right| \rho_{av}^{-1} =
\sum_{n=1}^{\infty} \left(
\frac{\varphi_{n}(\boldvec{x}_{1})}
{\omega-E_{n}} \right)^2 \rho_{av}^{-1} \nonumber \\
& \simeq & \left\langle \varphi_{n}(\boldvec{x}_{1})^{2} \right\rangle_n  
\sum_{n=1}^{\infty} 
\frac{2\rho_{av}^{-1}}
{\{(n-\frac{1}{2})\rho_{av}^{-1} \}^2} \nonumber \\
& = &   
\pi^{2} \left\langle \varphi_{n}(\boldvec{x}_{1})^{2} \right\rangle_n  
\rho_{av} = \frac{\pi M}{2}. 
\end{eqnarray}
The third equality follows from the approximation that 
the unperturbed eigenvalues are distributed with a mean 
interval $\rho_{av}^{-1}$ in the whole energy region. 

We can summarize the findings as follows; 
The effect of a point impurity of formal coupling strength $\bar{v}_1$ 
is substantial mainly in the eigenstates with eigenvalue $\omega$ 
such that 
\begin{eqnarray}
\label{eq:16}
\left| \bar{v}_1^{-1} - \frac{M}{2\pi} \ln \frac{\omega}{\Lambda}\right| \alt 
\frac{\pi M}{4}   
\end{eqnarray}
in two dimension. 
Numerical supports for the condition (\ref{eq:16}) are shown 
in \cite{SH94}--\cite{SM98}. 

\section{The Case of Finite-Size Impurity}

We have revealed the condition for the appearance of the effect 
of point impurities in the previous section. 
It should be noticed that the condition (\ref{eq:16}) 
is written in terms of the formal strength $\bar{v}_1$ as well as 
the scale mass $\Lambda$, either of which does not have a direct relation 
to the physical observables. 
It is realized, however, that both disappear in case of 
realistic finite-range impurities and the condition for 
the strong coupling can be described only in terms of the observables. 

Suppose that 
a small but finite-size impurity of the area $\Omega$ is 
located around $\boldvec{x}=\boldvec{x}_1$ 
inside the region $S$. 
We describe the impurity in terms of 
a potential which has a constant strength in the region $\Omega$; 
\begin{eqnarray}
\label{eq:17}
U(\boldvec{x}) = 
\left \{
\begin{array}{ll}
U_1, & \ \ \ \boldvec{x} \in \Omega, \\[1ex]
0  , & \ \ \ \boldvec{x} \in  S-\Omega.  
\end{array}
\right.
\end{eqnarray}
We assume that 
the impurity has the same order of size, say $R$, in each 
spatial direction, and also assume 
that the size of the impurity is substantially smaller 
than that of the outer region; 
$\Omega \simeq R^2 \ll S$. 
In this case, 
the impurity behaves as pointlike at low energy 
$\omega\ll E_{N(\Omega)}$, where $N(\Omega)$ is determined by 
$\displaystyle E_{N(\Omega)} \simeq 1/MR^2 \simeq 1/M\Omega$. 
Furthermore, the coupling of higher energy states 
than $E_{N(\Omega)}$ to the low-energy states 
is weak, since wave functions with wavelength shorter than 
$R$ oscillate within the impurity. 
This means that the low-energy states ($\omega \ll E_{N(\Omega)}$)  
can be described by the Hamiltonian (\ref{eq:4}) with the $\delta$-potential   
of the coupling strength 
$v_1 \equiv U_1  \Omega$,   
{\em together with a basis truncated at} $E_{N(\Omega)}$. 
The truncation of basis is crucial for the present argument. 
As mentioned before, in two dimension,  
the $\delta$-potential 
is not well-defined in the full unperturbed basis. 
The finiteness of the impurity introduces 
an ultra-violet cut-off in a natural manner and as a result,  
the low-energy dynamics can be reproduced by the Hamiltonian (\ref{eq:4}) 
within a suitably truncated basis. 

The strength $v_1$ can be related to 
the formal strength $\bar{v}_1$ as follows. 
Within the truncated basis $\left\{ \varphi_n (\boldvec{x}) \vert 
n = 1, 2, ..., N(\Omega)\right\}$, 
the eigenvalues of the Hamiltonian (\ref{eq:4}) are determined by 
\vspace{-1ex}
\begin{equation}
\label{eq:18}
\sum_{n=1}^{N(\Omega)} 
\frac{\varphi_{n}(\boldvec{x}_1)^2}{\omega-E_n} = v_1^{-1}.
\end{equation}
From (\ref{eq:9}), (\ref{eq:10}) and (\ref{eq:18}), 
we obtain 
\vspace{-1ex}
\begin{eqnarray}
\label{eq:19}
\lefteqn{
\bar{v}_1^{-1}  =  v_1^{-1} +   
\sum_{n=1}^{N(\Omega)} 
\varphi_{n}(\boldvec{x}_{1})^{2}\frac{E_{n}}{E_{n}^2+\Lambda^2} 
} \nonumber \\
&& + \hspace{-2ex} \sum_{n=N(\Omega)+1}^{\infty} \hspace{-2ex}
\varphi_{n}(\boldvec{x}_{1})^{2}\left(
\frac{1}{\omega-E_{n}}+\frac{E_{n}}{E_{n}^2+\Lambda^2}\right).   
\end{eqnarray}
The equation (\ref{eq:19}) gives an exact relation 
between $\bar{v}_1$ and $v_1$. 
In order to gain further insight on (\ref{eq:19}), 
we take the same approximation by integrals 
as in the previous section; 
\vspace{-2.5ex}
\begin{eqnarray}
\label{eq:20}
\bar{v}_1^{-1} & \! \simeq & \!  
v_1^{-1} +   \alpha 
\left\{ \int_{0}^{E_{N(\Omega)}} \frac{E}{E^2+\Lambda^2} dE 
\right. \nonumber \\
& \! & \! + \left. 
\int_{E_{N(\Omega)}}^{\infty} 
\left(\frac{1}{\omega-E}+\frac{E}{E^2+\Lambda^2}\right)dE \right\}. 
\end{eqnarray}
Inserting (\ref{eq:20}) into (\ref{eq:13}), 
we obtain the strong coupling condition 
for the finite-size impurity; 
\vspace{-0.5ex}
\begin{eqnarray}
\label{eq:21}
\left| v_1^{-1} - \alpha \cdot 
P  \int_{0}^{E_{N(\Omega)}} 
\frac{dE}{\omega-E} \right| 
\alt 
\frac{\Delta}{2}\simeq 
\frac{\pi M}{4}. 
\end{eqnarray}
This is exactly the condition for the eigenvalue equation 
(\ref{eq:18}) to have a solution $\omega$ around 
the midpoint on some interval $(E_m,E_{m+1})$. 
Performing the integration in (\ref{eq:21}) and noticing 
$v_1=U_1\Omega$, $\alpha \simeq M/2\pi$,  
we have the condition (\ref{eq:1}) for 
$\omega \ll E_{N(\Omega)} \simeq 1/M\Omega$. 
An arbitrary scale mass $\Lambda$ disappears and 
(\ref{eq:1}) is written in terms of the observables. 
This gives the general condition for the appearance of 
the effect of finite-size impurities on the electron motion 
in two-dimensional microscopic bounded regions;  
At low energy where finite-size impurities can be regarded 
as pointlike ($\omega \ll 1/M\Omega$), 
the electron (of effective mass $M$) 
is most strongly coupled to finite-size ($\simeq \Omega$) impurities   
of potential height $U_1$ under the condition (\ref{eq:1}). 

The most important indication of (\ref{eq:1}) is that 
the effect of finite-size impurities at low energy 
appears most strongly when it is weakly attractive. 
In order to confirm this numerically, 
we examine the wave function in a two-dimensional rectangular region with 
a small rectangular impurity inside \cite{SC96}. 
In the following, we set the scale mass $\Lambda=1$ without 
losing generality. 
The unperturbed eigenvalues and the corresponding normalized 
eigenfunctions without the impurity are given by 
\begin{equation}
\label{eq:22}
E_{m,n} = \frac{1}{2M}
\left\{ \left( \frac{m\pi}{l_x} \right)^2 +
        \left( \frac{n\pi}{l_y} \right)^2 \right\},
\end{equation}
and 
\begin{equation}
\label{eq:23}
\varphi_{m,n}(x,y) =
\sqrt{\frac{4}{l_x l_y}}
\sin \frac{m\pi x}{l_x}
\sin \frac{n\pi y}{l_y}, 
\end{equation}
with $m,n=1,2,3,\cdots$, respectively. 
We take the side-lengths of the (outer) rectangle as 
$(l_x,l_y) = (\pi/3,3/\pi)$. 
The mass of the particle is set to $M=2\pi$, 
leading to $\rho_{av}=1$. 
A small rectangular impurity with side-lengths 
$(\delta l_x, \delta l_y) = 
(3.53830 \times 10^{-2},  3.14023 \times 10^{-2})$ 
(area $\Omega=1/900$)
is placed at $\boldvec{x}_1=(0.622482,0.275835)$  
such that the sides of the inner and outer rectangles 
are parallel to each other. 
\begin{figure}
\begin{center}
\begin{tabular}{c}
\subfigure{\epsfile{file=rfig2b300,height=5cm}} \\[-17ex]
\subfigure{\epsfile{file=rfig2c300,height=5cm}} \\[-17ex]
\subfigure{\epsfile{file=rfig2d300,height=5cm}}  
\end{tabular}
\end{center}
\vspace*{-12ex}
\caption{Dependence of the eigenfunction on the nature of 
the impurity; 
(a) strong repulsion ($v_1 = 10$), 
(b) strong attraction ($v_1 = -3.33$), and  
(c) weak attraction ($v_1 = -0.25$). 
The eigenvalue $\omega$ of each state is 
(a) $\omega = 4.93$, 
(b) $\omega = 4.43$, and 
(c) $\omega = 5.63$, respectively.
The location of the impurity is denoted by a small rectangle. 
}
\label{fig:1}
\end{figure}

Fig.1 shows the dependence of the wave function on the nature of 
the impurity in the low-energy region. 
All the eigenstates are located between 
the unperturbed energies $E_{1,2}=4.16$ and $E_{2,2}=6.31$. 
Thus, the main components of each wave function are expected to be 
$\varphi_{1,2}$ and $\varphi_{2,2}$. 
In both cases of strong repulsion (a) and strong attraction (b), 
the wave function is dominated only by a single component 
$\varphi_{1,2}$ except around the small impurity, 
which is denoted by a small rectangle in Fig.1. 
It is worthy to note that 
the direct measure of the strength of the impurity is 
given by the ratio between $v_1=U_1\Omega$ and the mean level 
spacing $\rho_{av}^{-1}$. Since $\rho_{av}=1$, 
Both (a) and (b) indeed correspond to the strong force. 
Conversely, the mixture of the unperturbed eigenfunctions 
occurs in case of weak attraction (c), 
for which 
the strong coupling condition 
$1/U_1 M \Omega \simeq \ln(\omega M\Omega)/2\pi $ is satisfied 
with high degree of accuracy. 
These results confirm the validity of the prediction (\ref{eq:1}). 
For details, the readers are referred to \cite{SC96}. 

\section{Conclusion} 

We have discussed the effect of small impurities on the electron motion 
in two-dimensional microscopic bounded regions from a general perspective. 
The condition for the appearance of their effect is made clear 
in a quantitative manner. The equation (\ref{eq:1}) indicates 
the followings; 
\vspace*{0.5ex}
\begin{enumerate}
\item 
The effect of small impurities on the low-energy 
electron motion in two-dimensional 
microscopic bounded region appears when the potential is weakly 
attractive, while it can be neglected in case of strong force.  
\item 
The strong coupling region is described by 
a logarithmically energy-dependent strip with an energy-independent 
width in the $\omega$ versus $U_1^{-1}$ plane. This means that 
the strength of the small impurities which affect the electron dynamics  
changes as the electron energy increases. 
\end{enumerate}

\end{document}